\documentstyle[aps,epsf]{revtex}

\begin{document}
\title {Unified Description of Kaon Electroweak Form Factors\footnote
{\it Submitted to Physical Review C}}
\vskip 0.2 in
\author{Andrei Afanasev\thanks{Also at Kharkov Institute of Physics and
Technology, Kharkov 310108, Ukraine}\\ 
{\normalsize\it Thomas Jefferson National Accelerator Facility}\\ 
{\normalsize\it 12000 Jefferson Ave., 
Newport News, VA 23606, USA}\\
{\normalsize\it and}\\
{\normalsize\it The NuHEP Research Center, Department of Physics}\\ 
{\normalsize\it Hampton University, Hampton, 
VA 23668, USA}\\ W.W. Buck \thanks{Permanent Address: The NuHEP 
Research Center, Hampton University, Hampton, 
VA 23668, USA and TJNAF, 12000 Jefferson Ave., 
Newport News, VA 23606, USA}\\
 {\normalsize\it Institut f\"ur Kernphysik, Universit\"at Mainz, Germany}}
\maketitle

\begin{abstract}
A calculation of the semi--leptonic decays of the kaon ($K_{l3}$) 
is presented. The
results are direct predictions of a covariant model of the pion and kaon
introduced earlier by Ito, Buck, Gross. The weak form factors  
for $K_{l3}$ are predicted with absolutely no parameter adjustments
of the model. We obtained for the form factor parameters: 
$f_-(q^2=m_l^2)/f_+(q^2=m_l^2)=-0.28$  and $\lambda_+$= 0.028,
both within experimental error bars. Connections of this approach to heavy
quark symmetry will also be discussed.
\pacs\  PACS number(s): 11.10.St, 12.39.Ki, 13.20.Eb, 14.40.Aq
\end{abstract}


\section{Introduction}
The success of the Ito--Buck--Gross (IBG) \cite{ibg} model in the
description of many properties of both the $\pi$ and $K$ mesons
motivated the calculation of the $K_{l3}$ decays reported on here.
It is the  $K_{l3}$ decays that combine both the pion and kaon wave
functions generated previously [1-3]. A successful $K_{l3}$ calculation
that is coupled to other observables, constrains further the physics
described by the model.

The work reported on here is predictive and employs no new parameters and
no parameter adjustment. The results, found below, are in good agreement
with the data and are as good as Chiral Perturbation Theory (CPT) approaches
\cite{cpt} and \cite{shabalin},
even though a low energy parameter is involved in them, 
and better, at least in the light--quark sector, than the Quark Potential 
Model (ISGW2), with a hyperfine 
interaction,
predictions \cite{QM}. It is noted that an older version of the Quark 
Potential Model without
hyperfine interaction (ISGW) gives results similar to ours \cite{ISGW}. These 
comparisons and the details of our calculation are presented below. 
Nonetheless, with the success of CPT and this work, the question still 
remains for nuclear physics as to singling out quark structure from hadronic.
That is, where do hadrons leave off and quarks begin? The Quark Model has
been very successful at reproducing hadronic static properties such as the mass
spectrum and moments. But it is the dynamic properties, we feel, that will 
delineate
the differences between hadronic physics and quark physics. For this reason,
we take the position that not only is the near, or at $q^2=0$ physics
important but the $q^2\neq 0$ domain will delineate theoretical approaches.
Thus, our  predictions for the non--perturbative weak transition
form factors as a function of $q^2$
are also presented here in an attempt to draw both theoretical and empirical
interests. 

A detailed review of the theoretical 
and experimental status of semileptonic kaon decays is given in 
Ref.\cite{review}.

\section{The Model}

The theoretical model employed is an extension of the Nambu--Jona--Lasinio
(NJL) model but with a definite momentum distribution generated by a
separable interaction:
\begin{eqnarray} 
V(p,k)= g f(p^2) f(k^2) [I\otimes I-
(\gamma^5\tau)\otimes (\gamma^5\tau)],
\end{eqnarray}
where $f(p^2)=(\Lambda^2-p^2)^{-1}$ with $\Lambda$ being the interaction
cutoff parameter for a given meson state. With this choice of the 
$q\overline q$ interaction, one can integrate all momentum integrals to 
infinity; there is no need for an integral cutoff as employed in the NJL 
model.

The IBG model requires that the Bethe--Salpeter equation be solved for the 
vertex function, $\Gamma$, for each meson considered,


\bigskip

\epsfxsize=4 in
\hskip 2 cm \epsfbox{bsal.eps}

\bigskip

\noindent where $p$ is the 4--momentum of the meson.

The self energy of each flavour quark is treated by solving the 
Schwinger--Dyson equation
\begin{eqnarray}
\Sigma(k^2)= 4 i n_f g f(k^2)\int {d^4 p\over (2\pi)^4} f(p^2)
{m_0+\Sigma(p^2)\over p^2-[m_0+\Sigma(p^2)]^2}
\end{eqnarray}
where $n_f$ is the number of quark flavours (equal to 3 in our $q\overline q$ 
system)
in a coupled sense (coupled via quark masses and interaction strength) to the
Bethe--Salpeter equation. Though, in the case of the strange quark mass, the
self energy is assumed to be the (constituent) quark mass and is treated as
a parameter to be fixed. \cite{bwi}

Electomagnetic gauge invariance is imposed on the electromagnetic current
of a pseudoscalar meson
\begin{eqnarray}
J^\mu= F(q^2) (p+p')^\mu, 
\end{eqnarray}
where $F(q^2)$ is the meson charge form factor and $p (p')$ is the 4--momentum
of initial (final) meson. The work of Buck,
Williams, and Ito (BWI) \cite{bwi} has shown that both the pion and kaon
(charged and neutral) charge form factors can be predicted and it is the pion 
and kaon 
vertices from this work that are employed in the calculation of the weak form 
factors.

\section{ Weak Form Factors}

In the Standard Model, the weak current for $K_{l3}$ decays has the following 
structure
\begin{eqnarray}
J^\mu= {G_F\over \sqrt{2}} V_{us}[f_+(q^2)(P_K+P_\pi)^\mu+f_-
(q^2)(P_K-P_\pi)^\mu] 
\end{eqnarray}  
where $P_K$ and $P_\pi$ are the kaon and pion 4--momenta, $q=P_K-P_\pi$, 
and $f_\pm$ are dimensionless form factors. 

The semi--leptonic decays studied are:
\begin{eqnarray}
(K_{e3}) &K^\pm=\pi^0 e^\pm\nu_e \nonumber\\
 &K_L^0=\pi^\pm e^\mp\nu_e \nonumber\\ 
(K_{\mu 3}) &K^\pm=\pi^0 \mu^\pm\nu_\mu \\
&K_L^0=\pi^\pm \mu^\mp\nu_\mu \nonumber
\end{eqnarray}

In the limit of exact isospin symmetry, $m_u=m_d$, form factors of charged
and neutral kaon decays are related:
$$f_\pm^\pm/f_\pm^0=1/\sqrt{2},$$
and in the limit of exact SU(3) symmetry, the form factor $f_-$ is zero.
For the decay channel, the transferred 4--momentum $q$ is time--like, and
the physical region is limited to $m^2_l\leq q^2\leq (m_K-m_\pi)^2$.
The vertices appearing in this weak current and the ones employed in 
this work  are the kaon and pion vertices (wave functions) previously
obtained by BWI; namely
\begin{eqnarray}
\Gamma_{K,\pi}(k)= {N_{K,\pi} \gamma^5 \over\Lambda^2_{K,\pi}-k^2},
\end{eqnarray}
with $N_{K,\pi}$ being the normalization. 

From equation 4, one can uncouple $f_\pm$:
\begin{eqnarray}
f_\pm={(P_K\mp P_\pi)^2 J^\mu (P_K\pm P_\pi)^\mu-(m_K^2-m_\pi^2)
J^\mu (P_K\mp P_\pi)^\mu\over  (P_K-P_\pi)^2 (P_K+P_\pi)^2-
(m_K^2-m_\pi^2)^2}
\end{eqnarray}

To compare to available experimental data, the
following low-$q^2$ expansion is used for the form factors:
\begin{eqnarray}
f_\pm(q^2)=f_\pm(q^2=m_l^2) (1+\lambda_\pm {q^2-m_l^2\over m_{\pi^+}^2})
\end{eqnarray}
where $\lambda_\pm$ is the slope of $f_\pm$ evaluated at $q^2=m_l^2$
and $f_\pm(q^2=m_l^2)$ corresponds to the normalization. Note that it is
the charged, not the neutral, pion mass that enters the above expansion. 

Another set of the form factor parameters commonly used in the
literature  is $\lambda_+$, $\lambda_0$, arising as coefficients of
linear expansion of the form factors $f_+$ and $f_0$, with $f_0$
defined as
\begin{eqnarray}
f_0=f_++{q^2\over m_K^2-m_\pi^2} f_-,
\end{eqnarray}
The form factors $f_+$ and $f_0$ describe, respectively, $P-$wave and
$S-$wave projections of weak current matrix elements in the crossed channel.

To obtain
the values of $\lambda_\pm$ , a calculation of $J^\mu$ must be
performed. In this work, $J^\mu$ is the direct result of a triangle
diagram (Fig.1) with a flavour changing operator having $V-A$ spin structure
$\gamma^\mu (1-\gamma^5)$. In the Standard Model,
the $K_{l3}$ decay form factors are determined only by the vector
part of the charged weak current operator.

\begin{figure}
\let\picnaturalsize=N
\def\picsize{3in}
\def\picfilenamea{triangle.eps}
\ifx\nopictures Y\else{\ifx\epsfloaded Y\else\input epsf \fi
\let\epsfloaded=Y
\centerline{
\ifx\picnaturalsize N\epsfxsize \picsize\fi \epsfbox{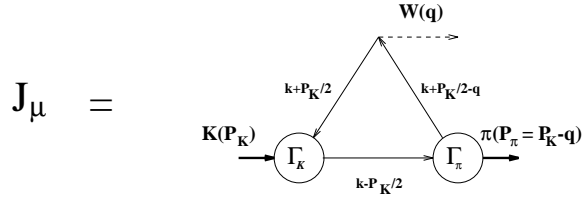}}}
\fi
\caption{Triangle diagram for the charged weak current of  $K_{l3}$ decay. }
\end{figure}
 
Integrals with respect to loop momentum were evaluated in the following
way: In the expession for the weak current given by a Feynman diagram Fig.1,
the spin trace was calculated and the terms dependent on loop momentum
in the numerator were divided out by corresponding terms in the denominator.
This procedure reduces the expression for `impulse' current Fig.1 to the
sum of scalar integrals of products of three to five denominator factors
(three of them coming from quark propagators and two from 
meson--$q\overline q$ vertex form factors). Each denominator factor 
is a polynomial
quadratic in the loop momentum. The terms involving three denominators,
are in fact scalar 3--point functions
which may be expressed analytically in terms of Spence functions \cite{tv}.
In this work, to calculate the 3--point functions, we parametrized them
in terms of two Feynman parameters. Integration with respect to one 
Feynman parameter was done analytically, and the other was a numerical
Gauss integration. We did not use a low $q^2$ expansion to evaluate loop
integrals but we do extract from our results the low $q^2$ behavior and find 
it is consistent with the low $q^2$ expansion of equation 8 above, 
employed by all researchers.

  Our task of computing 4-dimensional loop integrals of
products of more than three denominators is greatly simplified when
taking advantage of the fact that only two external momenta in the 
integrand are
linearly independent. As a result, products of four and five denominators
are reduced to the sum of products of three denominators with redefined
masses $M_i$. This procedure is described by the following identity
\begin{eqnarray}
{1\over{(k^2-m_0^2) ((k+q_1)^2-m_1^2) 
((k+q_2)^2-m_2^2) ((k+q_3)^2-m_3^2) ((k+q_4)^2-m_4^2)}}=\\
\sum_{i,j,n} {a_i\over
(k^2-M^2_i) ((k+q_j)^2-M^2_j) ((k+q_n)^2-M^2_n)},\nonumber
\end{eqnarray}
if
\[ \left( \begin{array}{c} q_3\\ q_4\end{array} \right)= {\bf A} 
\left( \begin{array}{c} q_1\\ q_2\end{array} \right),\ 
 {\rm det}{\bf A}\neq 0,\]
where the sum is taken over different combinations of the external
4--momenta $q_i$ involved in the reaction, $m_i$ are quark masses
and mass parameters in meson--$q\overline q$ vertex form factors, 
$a_i$ are coefficients independent of loop momentum $k$, and ${\bf A}$ is a 
2$\times$2 matrix setting relations between external momenta in the 
integrand. After this
reduction, scalar 3--point integrals are computed within the technique 
described above.

\section{Results}
In the physical region of $K_{e3}$ decays, $q^2$ may be as
low as the square of electron mass, 25$\cdot 10^{-8}$ GeV$^2$/c$^2$,
and as high as the square of the mass difference between the kaon and the
pion, 0.123  GeV$^2$/c$^2$. The form factors $f_\pm$ in this region 
with a good precision appear to be linear functions of $q^2$ thereby justifying
a linear parametrization of equation 8 usually employed in analyses of 
experimental 
data \cite{pdg}. To compare our results with experiment, we 
extracted the slopes and ratios of the form factors $f_\pm$ at    
$q^2=m_l^2$ via numerical differentiation.
Numerical values for the parameters in this calculation were taken
to be the same as in Ref.\cite{bwi}, $viz.$ $m_{u,d}$= 250 MeV,
$m_s$= 430 MeV, $\Lambda_\pi$= 600 MeV, and $\Lambda_K$= 690 MeV. 

The direct predictions of our approach for $\lambda_+$ and 
${f_-\over f_+}|_{q^2= m_l^2}$  are 0.028 and $-$0.28, respectively.
These results are to be compared to the experimental values of
$$\lambda_+=0.0286\pm 0.0022\ {\rm and}\ \xi_A=f_-/f_+=-0.35\pm 0.15.$$
 We obtain $\lambda_-$= 0.029, $i.e.$,
in our model both $f_-$ and $f_+$ have approximately the same slopes, in 
agreement with early Quark Model results \cite{isgur75}. 
Our calculation for $K_{e3}$ and $K_{\mu 3}$  yields equal results, within 
the quoted precision, since the $\lambda_\pm$ are almost constant in
the range $m_e^2\leq q^2\leq m_\mu^2$.
Naturally, the decay rates should
be different due to phase space factors; they can be calculated by known
formulas in terms for form factor slopes, see, $e.g.$ Ref.\cite{cpt}; 
however, we  
have yet to perform the calculation of these rates.

Table 1 illustrates the comparison between our work, that of CPT,
 vector meson dominance
(VMD), and the ISGW2 model. One sees that the work
reported on here compares
very favorably to experiment and to CPT, except for the prediction for
  $\lambda_0$.
\bigskip

Table1: Model predictions for the parameters of $K_{l3}$ 
decay form factors. 

$^*$ From the corresponding values of $\lambda_+$ and
$\lambda_0$ \cite{cpt}; $^{**}$ From the corresponding values of 
$\lambda_+$ and $\xi_A(0)$.
\medskip

\centerline{\begin{tabular}{|l|c|c|c|c|c|} \hline\hline
& CPT: Refs. \cite{cpt} and & VMD \cite{review} & ISGW2 \cite{QM}& This work&  Experiment
\cite{pdg} \\ 
& \cite{shabalin}, respectively& & & & \\ \hline
$\lambda_+$&0.031; 0.0328 & 0.0245 &
0.019 & 0.028& 0.0286$\pm$0.0022 ($K_{e3}$)\\ \hline
$\xi_A(0)$& --0.164$\pm$0.047$^*$;--0.235 & --0.28 
& --0.28&--0.28&--0.35$\pm$0.15 
($K_{\mu 3}$) \\ \hline
$\lambda_0$&0.017$\pm$0.004; 0.0128 & 0.0 
&
--0.005$^{**}$ & 0.0026& 0.004$\pm$0.007 ($K^+_{\mu 3}$) \\ 
& & & & & 0.025$\pm$0.006 ($K^0_{\mu 3}$) \\ \hline\hline
\end{tabular}}
\medskip

A prediction for the slope parameter $\lambda_0$ obtained within our model
is 0.0026, which is consistent with experiments on charged kaon decays 
($\lambda_0^+=0.004\pm 0.007$) and inconsistent with neutral kaon decay 
measurements
( $\lambda_0^0 =0.025\pm 0.006$). Since the experimental results for this
slope parameter are not firm, it is hard to draw any positive conclusions
about agreement or disagreement of our result for $\lambda_0$ with experiment.
However, we can compare it with predictions from other models. It can be seen 
from the Table 1 that
the quark model in general gives much smaller numbers for $\lambda_0$
than CPT.  

To test the sensitivity of our results, an arbitrary change in the 
$\Lambda_{\pi}$
cutoff from 600 MeV to 450 MeV (a 25\% change), results in a $\lambda_+$=
0.028 to 0.031 (an almost 10\% change), $\xi_A(0)$ from $-0.28$ to $-0.27$
($\approx$ 5\% change),
and  $\lambda_0$ from 0.0026 to 0.0058 ($\approx$ 100\% change), 
respectively. One is reminded that
changing $\Lambda_{\pi}$ changes the pion charge radius as well as the pion 
decay
constant. In fact, the value $\Lambda_{\pi}$= 450 MeV 
was used in Refs.[1,2] as 
the best fit to pion decay constant and charge radius alone, and this 
parameter was adjusted to 600 MeV in Ref.\cite{bwi} to be able to treat {\it 
both} pion 
{\it and} kaon in a coupled approach. 

It should be noted that our model
gives a stable, with respect to variation of the model parameter, prediction
for $\xi_A(0)$, and appears to give a highly parameter--dependent result for 
$\lambda_0$. This model dependence is due to cancellation between two large 
terms on the r.h.s. of equation (9). The situation is different in CPT, where 
uncertainties due to
higher--order loop corrections give rise to about 30\% uncertainty for
$\xi_A(0)$: $-0.164\pm 0.047$ and about $25\%$ uncertainty for $\lambda_0$
\cite{cpt}.
 
Finally, CPT suggests that once the ratio of the weak form factors is known, 
then an 
estimate of the mass of the strange sigma ($m_{\sigma_K}$), 
a meson with $J^P=0^+$, can be 
made. The relationship referred to is \cite{shabalin}
\begin{eqnarray}
\xi_A(0)=(M_K^2-M_\pi^2) (M_\rho^{-2} - M_{\sigma_K}^{-2}).
\end{eqnarray}

Taking our result for $\xi_A(0)$ and assuming duality between our model
predictions and the model with effective exchanges of vector and scalar
mesons at low $q^2$, we have  $m_{\sigma_K}$= 1.5 GeV, which compares
favorably to the mass of $K_0^*(1430)$.  
A test of this value could be made through the 
hypernuclear spectroscopy measurements (CEBAF E89-009; CEBAF PR--95--002) 
\cite{tang} inferring the interaction that contains this strange sigma.

Another feature of our approach is revealed in the limit as $\Lambda_K$,
$m_K$, and $m_s$ become infinitely large. The ratio $\xi_A(0)$ is calculated
in this limit and its asymptotic value is --1; Figure 2 illustrates the
mass dependence. In the limit of Heavy Quark Symmetry (HQS), the $q^2$ 
dependence
of the semileptonic decay form factors is factorized out in the form of
Isgur--Wise function \cite{hqs}, and $\xi_A$ is given by the combination
of the initial ($M$) and final ($M'$) meson masses,
\begin{eqnarray}
\xi_A=-{M-M'\over M+M'}.
\end{eqnarray}
As a result, $\xi_A|_{HQS}=-1$ if the initial meson is much heavier than the
final. Note that for HQS to be applied, both initial and final mesons should
be heavy, whereas assuming $m_s$ to be large in our model, we keep the final
meson light. This implies that this particular result of HQS appears to be more
generally applicable.
Of course, the ratio is zero for mesons of equal mass.

\begin{figure}
\let\picnaturalsize=N
\def\picsize{4.in}
\def\picfilenamea{xi.eps}
\ifx\nopictures Y\else{\ifx\epsfloaded Y\else\input epsf \fi
\let\epsfloaded=Y
\centerline{
\ifx\picnaturalsize N\epsfxsize \picsize\fi \epsfbox{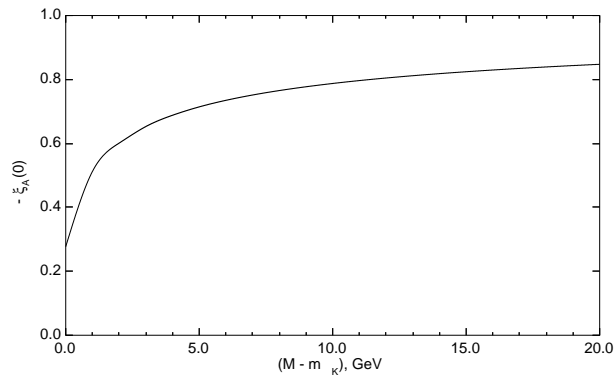}}}
\fi
\vskip -5 cm
\caption{Ratio $\xi_A(0)$ as a function of initial meson mass.}
\end{figure}

Though, it is tempting to make exuberant statements in regard to the 
identical results, one is cautioned by the manner in which the limits
are taken, and to the nature of the physics examined, respectively.

Furthermore, Figure 3 illuminates our predictions for $f_+(q^2)$ at
space--like momentum transfers describing the neutrinoproduction processes 
$\nu\pi\to lK$, $\nu\pi^0\to l\pi^+$ and corresponding weak lepton capture.
  We stress
that no low $q^2$ expansion was assumed in our calculations, so that
present results have the same validity range in terms of $q^2$ as results
of Refs.[1-3] for electromagnetic form factors of pion and kaon. The
form factor $f_+$ at large $q^2$ behaves as $1/q^4$ (up to a logarithmic
correction) indicating that our model effectively describes soft, 
nonperturbative reaction mechanism, and does not include perturbative QCD
contributions.

The weak $K\to\pi$ transition
form factors in the space--like region could be possibly accessed 
experimentally 
in the production of kaons
on an hadronic target induced by neutrinos or lepton weak capture.
The latter possibility is being studied for a CEBAF experiment \cite{finn}.

It would be instructive to see if the earlier success of the IBG model, that
include the pion and kaon observables as well as the results of this present
work, can be reproduced with other interactions and/or with other 
wave equations; by this, it is meant the predictive characteristics associated
with the low energy axial anomaly, such as the pion transition and elastic
charge  form factors, the kaon charge form factors, and the $K_{l3}$ decays.

In brief summary, weak form factors and slope parameters have been calculated
for $K_{l3}$ decays. The results compare very favourably to available 
experimental data. The model employed was that of IBG Refs.[1-3] and there
were no parameter adjustments, thus, rendering this calculation predictive.


\acknowledgements
The work of A.A. was supported by the US Department of Energy under
contract DE--AC05--84ER40150; the work of W.W.B. was supported by the 
National Science Foundation Grant Number HRD-9154080.
We would like to acknowledge useful discussions with J. Goity, N. Isgur,\ \
A. Radyushkin, and R. Williams in the course of this work.


\begin{figure}[h]
\let\picnaturalsize=N
\def\picsize{5.in}
\def\picfilenamea{fq2.eps}
\ifx\nopictures Y\else{\ifx\epsfloaded Y\else\input epsf \fi
\let\epsfloaded=Y
\centerline{
\ifx\picnaturalsize N\epsfxsize \picsize\fi \epsfbox{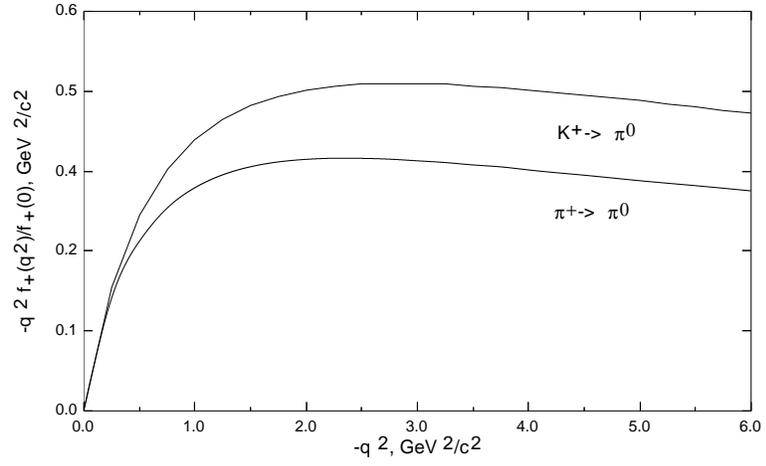}}}
\fi
\vskip -6 cm 
\caption{Form factors of weak transitions $K^+\to\pi^0$ and $\pi^+\to \pi^0$
at space--like transferred momenta.}
\end{figure}

\end{document}